\documentclass[aps,prb,preprint,showpacs,floatfix,superscriptaddress]{revtex4-1}
\usepackage{graphicx,bm,amsmath,dcolumn}
\begin{document}
\newcommand{\be}{\begin{equation}}
\newcommand{\ee}{\end{equation}}
\newcommand{\bea}{\begin{eqnarray}}
\newcommand{\eea}{\end{eqnarray}}
\newcommand{\half}{\frac{1}{2}}
\newcommand{\ith}{^{(i)}}
\newcommand{\im}{^{(i-1)}}
\newcommand{\gae}
{\,\hbox{\lower0.5ex\hbox{$\sim$}\llap{\raise0.5ex\hbox{$>$}}}\,}
\newcommand{\lae}
{\,\hbox{\lower0.5ex\hbox{$\sim$}\llap{\raise0.5ex\hbox{$<$}}}\,}
\newcommand{\mat}[1]{{\bf #1}}

\title{Ising-like transitions in the O($n$) loop model on the square lattice}
\author{Zhe Fu}
\affiliation{Physics Department, Beijing Normal University,
Beijing 100875, P. R. China}
\author{Wenan Guo} 
\email{waguo@bnu.edu.cn}
\affiliation{Physics Department, Beijing Normal University,
Beijing 100875, P. R. China}
\author{Henk W. J. Bl\"ote} 
\email{henk@lorentz.leidenuniv.nl}
\affiliation{Instituut Lorentz, Leiden University,
P.O. Box 9506, 2300 RA Leiden, The Netherlands}
\date{\today} 
\begin{abstract}
We explore the phase diagram of the O($n$) loop model on the square 
lattice in the  $(x,n)$ plane, where $x$ is the weight of a lattice
edge covered by a loop.
These results are based on transfer-matrix calculations and finite-size
scaling. We express the correlation length associated with the staggered
loop density in the transfer-matrix eigenvalues. The finite-size data
for this correlation length, combined with the scaling formula,
reveal the location of critical lines in the diagram.
For $n>>2$ we find Ising-like phase transitions associated with the
onset of a checkerboard-like ordering of the elementary loops,  i.e.,
the smallest possible loops, with the size of an elementary face, which
cover precisely one half of the faces of the square lattice at the
maximum loop density. In this respect, the ordered state resembles that
of the hard-square lattice gas with nearest-neighbor exclusion, and the
finiteness of $n$ represents a softening of its particle-particle potentials.
We also determine critical points in the range $-2\leq n\leq 2$.
It is found that the topology of the phase diagram depends on the set of
allowed vertices of the loop model. Depending on the choice of this set,
the $n>2$ transition may continue into the dense phase of the $n \leq 2$
loop model, or continue as a line of $n \leq 2$ O($n$) multicritical 
points.
\end{abstract}
\maketitle 

\section{Introduction}
\label{intro}
\begin{figure}
\includegraphics[scale=0.40]{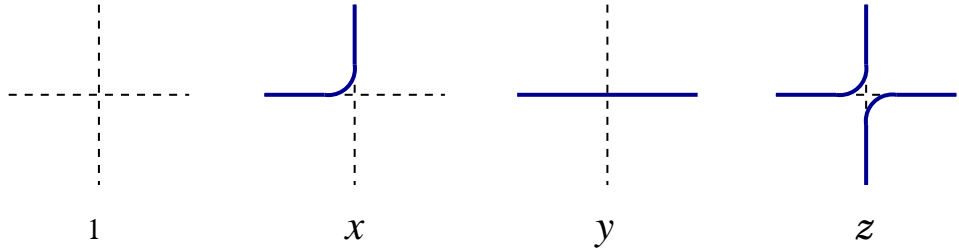}
\centering
\caption{(Color online) 
The four kinds of vertices of the O($n$) loop model on the square lattice,
together with their weights. Rotated versions have the same weights.
The present work is restricted to two subspaces of $(x,y,z)$, namely
$y=x$, $z=x^2$ and $y=0$, $z=x^2$.
\\}
\label{vertices}
\end{figure}

The O($n$) loop model is a highly useful tool for the analysis of O($n$)
symmetric $n$-component spin models \cite{Stanley,Domea}, and also for 
that of polymers \cite{deG,Np,DS}. A number of such loop models in two
dimensions is exactly solvable \cite{N,Baxter,BNW,3WBN,3WPSN,VF,S,PS,GNB}.
The present work investigates the nonintersecting loop model described by
the partition sum
\begin{equation}
Z_{{\rm loop}} =\sum_{{\rm all} \; {\mathcal G}}
x^{N_x} y^{N_y} z^{N_z} n^{N_l}, 
\label{Zloop}
\end{equation}
where ${\mathcal G}$ is a graph consisting of any number of $N_l$ closed,
nonintersecting loops. 
Each lattice edge may be covered by at most one loop segment, and  
there can be 0, 2 , or 4 incoming loop segments at a vertex. In the
latter case, they can be connected in two different ways without having 
intersections. The allowed four kinds of vertices configurations are
shown in Fig.~\ref{vertices}, together with their weights denoted $x$,
$y$ and $z$.  The numbers of vertices with these weights are denoted
$N_x$, $N_y$, $N_z$ respectively.

This loop model is equivalent with an O($n$) spin model, as described by 
Ref.~\cite{BN}. The $n$-component spins are sitting in the middle
of the edges of the square lattice. The Boltzmann weight for each spin
configuration is the product over
all vertices of the lattice of the local weights $w$
\bea
w &=&1+ x (\vec{s}_{1} \cdot \vec{s}_{2}+
\vec{s}_{2} \cdot \vec{s}_{3}+ \vec{s}_3 \cdot \vec{s}_4 + 
\vec{s}_{4} \cdot \vec{s}_{1}) +
y (\vec{s}_{1} \cdot \vec{s}_{3}+\vec{s}_{2} \cdot \vec{s}_{4}) \nonumber \\
& &+z [(\vec{s}_{1} \cdot \vec{s}_{2})(\vec{s}_{3} \cdot \vec{s}_{4})+
(\vec{s}_{2} \cdot \vec{s}_{3})(\vec{s}_{4} \cdot \vec{s}_{1})],
\label{Zspin}
\eea
where the spins $\vec{s}_1$ to $\vec{s}_4$ sit on the four edges incident
to the vertex and are labeled anticlockwise. 
The spins are subject to a measure and normalization
\begin{equation}
\int d{\vec{s}}_k=1\,,~~~~~\int d{\vec{s}}_k(\vec{s}_{k}\cdot\vec{s}_{k})=n\,.
\end{equation}
Expansion of the partition integral in powers of the coupling constants
$x,y$ and $z$ turns the spin model into the loop model of Eq.~(\ref{Zloop}).

Although the spin dimensionality $n$ can assume only integer values
$n=1,~2,~\cdots$ in the original O($n$) spin model,
$n$ can also have noninteger and even negative values in the loop
model of Eq.~(\ref{Zloop}), while the partition sum remains well-defined.
Whereas the Boltzmann weight of Eq.~(\ref{Zspin}) can become negative
when $x$, $y$ and/or $\sqrt z$ exceeds values of order $1/n$, 
the Boltzmann weights of the loop model remain 
physical for all non-negative values of $x$, $y$, $z$ and $n$.

The parameter space of Eq.~(\ref{Zloop}) contains several exactly solved
``branches'' \cite{BNW,3WBN}. These solutions have shown the existence of
a richness of phase behavior and ``nonuniversal'' lines as a function of
the vertex weights and $n$ in the range $n<2$. Branches 1 and 2 describe
the universal properties of the O($n$) critical point and the low-temperature
dense loop phase. Branch 4 describes the superposition of the low-temperature
phase 
and an Ising-like transition where a twofold symmetry of the loop
configurations is broken, and branch 3 describes the multicritical point
where the Ising-like line in the $(x,z)$ (for constant $y/x$) diagram
merges with the O($n$)  critical line \cite{BN}.
One can visualize the Ising-like degrees of freedom by assigning $+$ and
$-$ spins to the faces of the square lattice, according to the rule that
nearest-neighboring spins are equal only if there is a loop segment in between.

Our present aim is to supplement these findings with an analysis of
the model of Eq.~(\ref{Zloop}), which has not been solved for general $n$,
and with particular attention to the range $n>2$ where we expect an
Ising-like transition line. This expectation is based on the observation
that, in the limit of large $n$, the local weights are maximal for
configurations with small loops on the elementary faces of the lattice.
At most one half of these faces can be covered by a loop. There exist
two checkerboard-like configurations at maximum covering, similar to the
ordered phase of the hard-square model with nearest-neighbor exclusion.
We may thus expect a transition for sufficiently large $x$ in the Ising
universality class. Since we have set $z=x^2$, there is, except for the
nearest-neighbor exclusion, no further interaction between the hard
squares in the limit $n \to \infty$.

The present work will focus on two subspaces parametrized by $n$ and
the bond weight $x$, with $y=x, z=x^2$ and $y=0, z=x^2$.
For large $n$, we expect only small loops, and similar behavior in
both cases. However, for small $n$ larger loops exist with, if
$y \ne 0$, straight segments due to the $y$-type vertices. As
explained in Ref.~\cite{BN}, these  $y$-type vertices are responsible
for the flipping of an Ising-like degree of freedom along the loops. 
Thus, for small $n$ we may expect qualitative differences between
the cases  $y=x$ and $y=0$.

In Sec.~\ref{transfmat} we sketch our numerical procedures used to
locate the phase transition lines in the phase diagram.
Section \ref{results} presents the analysis of these
numerical results, as well as the resulting phase diagram.
The conclusions are summarized and discussed in Sec.~\ref{disc}.

\section{The transfer-matrix analysis}
\label{transfmat}

Our analysis is based on the numerical transfer matrix (TM) calculation
of $Z_{\rm loop}$ for $L\times \infty$ square lattices  wrapped on a
cylinder with circumference $L$. The transfer matrix keeps track of the
change of the numbers of loops and the four kinds of vertices when a
new layer of $L$ sites is added. The TM techniques for the O($n$) loop
model and the procedure for the sparse-matrix decomposition are already
described in the literature, e.g., see Ref.~\cite{BN}. 

The largest eigenvalue $\Lambda_0$ of the TM determines the free energy 
density $f(L)$ by
\begin{equation}
f(L)=\frac{\ln(\Lambda_0)}{L} \, .
\end{equation}
Its finite-size-scaling behavior at the critical point determines the
conformal anomaly $c$ according to \cite{BCN,Aff}
\begin{equation}
f(L) \simeq f(\infty)+\frac{\pi c}{6 L^2}+\cdots \, .
\label{fc}
\end{equation}

The magnetic correlation function of the O($n$) spin model over a distance
$r$ can be expressed in terms of the probability that two vertices at this
distance are connected by a single loop segment \cite{CG}. Thus one may write 
\begin{equation}
g_{m}(r)=\frac{Z'}{Z} \, ,
\end{equation}
where $Z'$ is the same as in Eq.~(\ref{Zloop}), but with the sum on all loop
configurations ${\mathcal G}'$ that contain one additional single loop
segment that runs from  position 0 to $r$. 

The exponential decay of $g_{m}(r)$ at large distances is determined by
the  magnetic correlation length $\xi_h(x,n,L)$, which can be obtained
numerically as
\begin{equation}
\xi_h^{-1}(x,n,L)= \ln \left(\frac{\Lambda_0}{\Lambda_1} \right) \, ,
\label{xih}
\end{equation}
where $\Lambda_1$ is the largest eigenvalues in the ``magnetic sector'',
which refers to the TM for $Z'$, which is based on loop configurations
with an additional single loop segment running along the cylinder.
The scaled magnetic gap $X_h$ is defined as
\begin{equation}
X_h(x,n,L)= \frac{L}{2 \pi \xi_h(x,n,L)} \, .
\label{xh}
\end{equation} 
Its finite-size-scaling behavior \cite{JCxi,FSS} near a critical point
$x_c$ is given by
\begin{equation}
X_h(x,n,L)= X_h + a (x-x_c) L^{y_t} + b L^{y_u} + \cdots  \, ,
\label{xhse}
\end{equation}
where $X_h$ is the magnetic scaling dimension, $y_t=2-X_t$ the
temperature exponent, and $y_u$ the leading irrelevant exponent.
The amplitudes $a$ and $b$ are nonuniversal quantities.  

In general, one expects that two phase transitions may occur in the
two-dimensional O($n$) model with $n\le 2$ on the square lattice when
the bond weight is increased \cite{BN}. The first one is the transition
from the dilute loop phase to the low-temperature O($n$) phase,
where the loops are densely packed. The size of the longest loop
diverges at this transition point. The universal properties
of this transition follow from the exact solution \cite{BNW,3WBN}
for branch 1, and from the Coulomb gas analysis \cite{CG}.
The results for the conformal anomaly and the magnetic exponent are 
\bea
c&=&1-\frac{6(g-1)^2}{g} \, ,\nonumber \\
X_h&=&1-\frac{1}{2g}-\frac{3g}{8} \, ,
\label{cXh}
\eea
where $g$ is the Coulomb gas coupling, which is related to $n$ by 
$n=-2\cos(\pi g)$ and $1 \le g \le 2$.  
The low-temperature phase is still critical in the sense that the magnetic
correlation function decays algebraically in the infinite system. 
The universal properties of the low-temperature phase are characterized
by a conformal anomaly $c^{({\rm LT})}$ and a magnetic scaling dimension 
$X_h^{({\rm LT})}$, which can be obtained from the results \cite{BNW,3WBN,CG}
for  branch 2 of the square-lattice loop model.  They are still given 
by Eq.~(\ref{cXh}) and $n=-2\cos(\pi g)$, but with $0\le g \le 1$. 

A second transition may occur inside the low-temperature phase, when the
loops enter an even denser phase which breaks the Ising-like symmetry of
the loop configurations \cite{BN}. Its universal properties \cite{BN,BNW}
were derived from the solvable case denoted as branch 4.
The magnetic scaling dimension $X_h^{({\rm LTI})}$ and the conformal anomaly 
$c^{({\rm LTI})}$ at this Ising-like transition correspond with a combination
of low-temperature O($n$)  and Ising-like critical behavior, namely
\be
X_h^{({\rm LTI})}=X_h^{({\rm LT})}+1/8
\label{Isingxh}
\ee
and
\be 
c^{({\rm LTI})}=c^{({\rm LT})}+1/2 \, .
\label{Isingc}
\ee

To analyze the expected transition for $n>2$, which drives the loop
gas into a loop ``solid'' phase with a checkerboard pattern,
we introduce  the staggered loop density and interpret it as the order
parameter. First, we define a face as ``occupied'' by a loop if it is
surrounded by a loop or any odd number of loops. In analogy with the
hard-square lattice gas, we also divide the faces of the lattice into
``odd'' and ``even'' ones.  Then one defines the staggered loop density
as the density of the occupied odd faces minus that for the even faces.
The staggered lattice gas correlation function is thus
\begin{equation}
g_s(r)=\langle \rho_s(0) \rho_s(r) \rangle \, ,
\end{equation}
where $\rho_s(0)$ and $\rho_s(r)$ are the staggered densities at 
positions $0$ and $r$, respectively.  
For large $n$, we expect that the dense phase is dominated by configurations
of elementary loops covering either the even or the odd faces.
Therefore, the staggered correlation function is associated with the
leading eigenvector $\vec{v}$ of the TM that is antisymmetric under the
operation ${\bf R}$ i.e., 
\be
\vec{v}=-{\bf R} \vec{v} \, ,
\label{steiv}
\ee
where ${\bf R}$ is the operator that rotates the lattice
by one lattice unit about the axis of the cylinder. 
As a consequence of the Perron-Frobenius theorem, the absolute value of
the corresponding TM eigenvalue $\Lambda_2$ cannot exceed $\Lambda_0$
which is associated with a symmetric eigenvector, at least for $n>0$.
We expect that the staggered correlation function scales in a similar
way as the magnetic correlation function. Thus we describe the exponential
decay of the staggered correlation function along the cylinder by means
of the staggered scaled gap, defined as 
\begin{equation}
X_s(x,n,L)=\frac{L}{2 \pi} \ln \left(\frac{\Lambda_0}{\Lambda_2}\right)\, .
\end{equation}

The scaled gap $X_s(x,n,L)$ is expected to behave according to  
Eq.~(\ref{xhse}), with $X_h$ replaced by the staggered lattice gas
scaling dimension $X_s$.
This transition breaks the $Z_2$ symmetry of odd and even lattice faces, 
and is thus expected in the Ising universality class: $c=1/2$ and $X_s=1/8$.

The critical point can be estimated by numerically solving $x$ in the
scaling equation involving two different system sizes 
\begin{equation}
X_i(x,n,L)=X_i(x,n,L'),\ \ (i=h,s) \, ,
\label{xhs}
\end{equation}
of which the solution $x_c(L)$ scales as 
\begin{equation}
x_c(L)=x_c+a' L^{y_u-y_t}+\cdots \, ,
\label{xcs}
\end{equation} 
where $a'$ is an unknown constant. 
Because $y_{u}<0$ and  $y_{t}>0$,
$x_c(L)$ converges to the critical point $x_c$ for a sequence of
increasing system sizes $L$.
At  $x_c(L)$, the scaled gap in Eq.~(\ref{xhse}) 
converges to the magnetic or the staggered 
lattice gas scaling dimension $X_i(n)$ according to the scaling equation
\begin{equation}
X_i(x_c(L),n,L)=X_i(n)+b'  L^{y_u}+\cdots \, ,
\label{xis}
\end{equation}
with an unknown amplitude $b'$.
An alternative way to obtain estimates $x_c(L)$ of the critical point 
is to neglect the correction term  and thus to solve for $x$ in the
equation
\begin{equation}
X_i(x,n,L)=X_i(n) \, ,
\label{xhs2}
\end{equation}
where $X_i(n)$ is the theoretical prediction for the pertinent scaling
dimension. Such predictions can follow the assumption that Eq.~(\ref{cXh})
or the Ising magnetic scaling dimension $1/8$ applies. If this assumption
is correct, the solutions of Eq.~(\ref{xhs2}) converge to the critical
point $x_c$ as described by Eq.~(\ref{xcs}). If the assumption is not
correct, then the finite-size dependence of the solutions behaves as
$L^{-y_t}$, so that they still converge to the critical point for
$y_t>0$, but relatively slowly. The finite-size dependence of the 
solutions of Eq.~(\ref{xhs2}) may thus reveal if the assumed value of
$X_h$ is right.

In the present work, we shall make use of both Eq.~(\ref{xhs})
and Eq.~(\ref{xhs2}) to determine the critical points. In most of these
calculations we restrict the system size to even values, because dense
loop configurations do not fit well in odd systems, which thus leads to
an even-odd alternation effect.

\section{Results}
\label{results}
In the range $-2 \le n \le 2$, much is already known about the general
properties of the phase diagram of the O($n$) loop model on the square
lattice \cite{BNW,BN,3WBN,GBN,GB}. 
This is not the case for the range $n>2$.
In this section we explore the phase diagram as a function of $n$, for
two types of loop model described by the single bond weight $x$.

\subsection{The subspace $y=x$, $z=x^2$}
In this subsection we explore the phase diagram for the case that there are
no  further conditions on the set of allowed vertices, thus with vertex
weights $x$, $y=x$ and $z=x^2$. 

For $n \le 2$, we estimate $x_c$ for the O($n$) critical and the
low-temperature branches 
by extrapolating the solutions $x_c(L)$ of  Eq.~(\ref{xhs2}) 
with $X_h$ the magnetic scaling dimension of the O($n$) critical branch
and the low-temperature branch, respectively. This procedure still
leads to convergent results in the case that the temperature field 
associated with $x$ is irrelevant, but less irrelevant than the other
nonzero scaling fields, i.e., the case expected for the LT phase of
the O($n$) model with not too small $n$, on the basis of the results for
branch 2 \cite{3WBN,BN}.

The transfer-matrix calculations were performed for system sizes up to
$L=16$ or, in some cases, 18. 
We then calculated the free energy density $f(L)$ at the estimated $x_c$,
and obtained the conformal anomaly $c$ by fitting these data according
to Eq.~(\ref{fc}). The numerical results for $x_c$ and $c$ are listed
in Tables \ref{yne0nlt2} and \ref{yne0nlt2b} respectively.  
Our numerical estimations of $c$ agree well with the theoretical
predictions, except for $n=-2$ where the finite-size data  display
poor convergence.

\begin{table}[tpbd]
\caption{Numerical results for the critical points $x_c(n)$ and conformal 
anomaly $c(n)$ for the O($n$) critical branch in the range $-2\le n \le2$ 
in the subspace $y=x, z=x^2$. 
The theoretical values for branch 1 of the conformal anomaly are also listed.
Estimated numerical uncertainties in the last decimal place are shown
between parentheses.}
\begin{tabular}{|c|c|c|c|}
\hline
 $n$     & $x_c(n)$          &$c_{\rm num}$   &$c_{\rm br1}$ \\
\hline
$-$2     & 0.33732317(2)     &$-$1.69(1)      &$-$2          \\
$-$1.5   & 0.3444544(1)      &$-$1.009(1)     &$-$1.00961    \\
$-$1.0   & 0.35259515(1)     &$-$0.60000(2)   &$-$3/5        \\
$-$0.5   & 0.3620756(1)      &$-$0.27901(1)   &$-$0.279017   \\
   0.0   & 0.3734237(1)      &   0            &   0          \\
   0.5   & 0.38757234(1)     &   0.25594(1)   &   0.255949876\\
   1.0   & 0.406446(2)       &   0.50000(1)   &   1/2        \\
   1.5   & 0.4353496(1)      &   0.74183(2)   &   0.74184247 \\
   1.8   & 0.4664502(1)      &   0.89185(1)   &   0.89185788 \\
   1.9   & 0.484688(1)       &   0.94432(1)   &   0.9443219  \\
   1.95  & 0.4988697(1)      &   0.97151(1)   &   0.971508   \\
   1.98  & 0.512488(1)       &   0.98835(1)   &   0.988346   \\
   1.99  & 0.519766(2)       &   0.99411(2)   &   0.994103   \\
   2.0   & 0.5386256(2)      &   1.00000(1)   &   1          \\
\hline
\end{tabular}
\label{yne0nlt2}
\end{table}

\begin{table}[tpbd]
\caption{Numerical results for the critical point $x_c(n)$ and the conformal 
anomaly $c(n)$ for the low temperature branch (branch 2) in the range
$-2\le n \le2$ in the subspace $y=x, z=x^2$. 
The theoretical predictions of conformal anomaly are also listed.
Estimated numerical uncertainties in the last decimal place are shown
between parentheses.}
\begin{tabular}{|c|c|c|c|}
\hline
 $n$     & $x_c(n)$       & $c$ (numerical) & $c$ (theory)\\
\hline
1.99     & 0.559583(1)    & 0.99372(2)      & 0.993716   \\
1.98     & 0.568975(1)    & 0.98725(2)      & 0.987247   \\
1.95     & 0.58904(1)     & 0.96714(3)      & 0.967132   \\
1.9      & 0.6145(1)      & 0.93180(1)      & 0.93179998 \\
1.8      & 0.657(1)       & 0.8557(1)       & 0.855602   \\
1.5      & 0.72(1)        & 0.588(1)        & 0.587572   \\
1.4      & 0.74(1)        & 0.485(1)        & 0.4849998  \\
\hline
\end{tabular}
\label{yne0nlt2b}
\end{table}

For the Ising-like transition in the LT dense phase, 
the numerical results for $x_c$ at $n=2$ and $n=1.5$  are extrapolated
from the solutions of the finite-size scaling equation (\ref{xhs}) for
$X_h(L)$ with even $L$.
For other values of $n$, the critical points $x_c$ are extrapolated from
the solution of the scaling equation 
(\ref{xhs2}), in which $X_h$ is taken as $X_h^{({\rm LTI})}$,
for even system sizes up to $L=16$. 
The free energy density $f(L)$ at the estimated critical point is then
calculated for even system sizes up to $L=16$.
A fit of these data thus yields the conformal anomaly $c$,
in a good agreement with $c^{({\rm LTI})}$ given in Eq.~(\ref{Isingc}).
The numerical results for $x_c$ and $c$ are listed in Table \ref{numdxlt}.
\begin{table}[tbpd]
\caption{Numerical results for the critical points $x_c(n)$ and the
conformal anomaly $c(n)$ for the Ising-like transition inside the LT
dense phase in the subspace $y=x, z=x^2$. Estimated numerical
uncertainties in the last decimal place are shown in parentheses. 
The results for $c(n)$ agree well with the 
with the theoretical predictions $c_{\rm br4}$ for branch 4 of the
square O($n$) model.
}
\begin{tabular}{|r|l|l|l|}
\hline
 $n$ ~~~~   & ~~~$x_c(n)$     & ~~~$c(n)$       &~~~$c_{\rm br4}$\\
\hline
$-$1.5 ~    & 1.580(2)        &$-$14.4(4)       &$-$13.9612 \\
$-$1.0 ~    & 1.463(1)        &$-$6.51(1)       &$-$6.5     \\
$-$0.5 ~    & 1.38398(1)      &$-$3.318(1)      &$-$3.31779 \\
~~ 0.0 ~    & 1.3229(1)       &$-$1.501(1)      &$-$1.5     \\
~~ 0.5 ~    & 1.27287(2)      &$-$0.318(3)      &$-$0.319736\\
~~ 1.0 ~    & 1.23019(2)      &~~ 0.50000(4)    &~~ 0.5     \\
~~ 1.2 ~    & 1.21468(1)      &~~ 0.759(2)      &~~ 0.758346\\
~~ 1.5 ~    & 1.19282(2)      &~~ 1.088(1)      &~~ 1.08757 \\
~~ 2.0 ~    & 1.15943(2)      &~~ 1.502(2)      &~~ 1.5     \\
\hline
\end{tabular}
\label{numdxlt}
\end{table}

Next, we explore the phase diagram for $n>2$.  For relatively large values
$(n\ge 8)$, we numerically solved for $x_c(L)$ in Eq.~(\ref{xhs2}) with
the expected Ising value $X_s=1/8$.
For $n=3,4,5$, we solved for $x_c(L)$ in the scaling equation (\ref{xhs}).
The critical point $x_c$ is then estimated for several values of $n$,
according to the scaling behavior given in Eq.~(\ref{xcs}).
We then calculated the free energy density $f(L)$ at the estimated 
critical points for even system sizes up to $L=18$.
>From a fit of Eq.~(\ref{fc}) to the data, we can thus obtain 
the conformal anomaly $c$. The best estimates  of $x_c$ and $c$ are listed
in Table \ref{numdx} for several values of $n>2$. Our estimates of $c$
agree well with the expected value $c=1/2$ for the Ising universality
class, at least for large $n$.  
However, for smaller values of $n$, the estimates of $c$ deviate from 1/2. 
This may be attributed to strong correction to scaling associated with the
expected marginal temperature field at $n=2$.

In the limit $n \to \infty$, only loops of the smallest possible size occur,
with the size of an elementary square of the lattice. The model of
Eq.~(\ref{Zloop}) then reduces to the lattice gas on the square lattice
with nearest-neighbor exclusion and no further interactions. We make use of
the existing numerical result $\mu_c=1.3340151002774(1)$ for the critical
value of the chemical potential of this model \cite{GBhsq}. By relating
the weight of an elementary loop to this chemical potential, which leads
to $nx_c^4=\exp (\mu_c)$, we obtain the large-$n$ limiting behavior of $x_c$.

\begin{table}[tbpd]
\caption{Numerical results for the critical points $x_c(n)$ and the
conformal anomaly $c(n)$ for the lattice-gas-like phase transitions.
Results are shown for the two cases $y=x$ and $y=0$.
Estimated numerical uncertainties in the last decimal place are shown
in parentheses.  The results for $c$ agree well with the expected Ising
conformal anomaly $c=1/2$, except for $n=2$ and for some relatively
small values of $n$ where poor convergence occurs.}
\begin{tabular}{|r||c|c||c|c|}
\hline
&\multicolumn{2}{|c||}{case $y=x$}&\multicolumn{2}{c|}{case $y=0$} \\
\hline
$n$   & $x_c(n)$   & $c(n)$   & $x_c(n)$     & $c(n)$     \\
\hline
 2    & --------   & -------- & 0.784(1)     & 1.49(2)    \\
 3    & 1.101(2)   & 2.0(6)   & 0.809(1)     & 1.3(2)     \\
 4    & 1.12(2)    & 1.5(2)   & 0.800(4)     & 0.9(3)     \\
 5    & 1.00(2)    & 1.3(2)   & 0.787(3)     & 0.6(3)     \\
 8    & 0.907(2)   & 0.5(1)   & 0.746(1)     & 0.50(1)    \\
10    & 0.857(3)   & 0.51(1)  & 0.7202(1)    & 0.50(1)    \\
15    & 0.7716(3)  & 0.501(3) & 0.6696(1)    & 0.50(1)    \\
20    & 0.7147(3)  & 0.500(1) & 0.6322(1)    & 0.50(1)    \\
30    & 0.64060(1) & 0.500(1) & 0.5795(2)    & 0.498(2)   \\
40    & 0.59240(1) & 0.500(1) & 0.54318(1)   & 0.500(2)   \\
50    & 0.55754(1) & 0.500(1) & 0.51593(1)   & 0.500(2)   \\
75    & 0.4995(1)  & 0.500(1) & 0.46888(1)   & 0.500(1)   \\
100   & 0.4622(2)  & 0.500(1) & 0.43760(1)   & 0.500(1)   \\
200   & 0.3841(1)  & 0.500(1) & 0.36958(1)   & 0.500(1)   \\
400   & 0.32004(1) & 0.5001(1)& 0.311448(1)  & 0.5000(1)  \\
800   & 0.26726(1) & 0.5001(1)& 0.262179(2)  & 0.5000(1)  \\
1000  & 0.25230(1) & 0.5001(1)& 0.248007(1)  & 0.5000(1)  \\
10000 & 0.14033(1) & 0.5000(1)& 0.139573(1)  & 0.50000(1) \\
\hline
\end{tabular}
\label{numdx}
\end{table}

\begin{figure}[tbpd]
\includegraphics[scale=0.8]{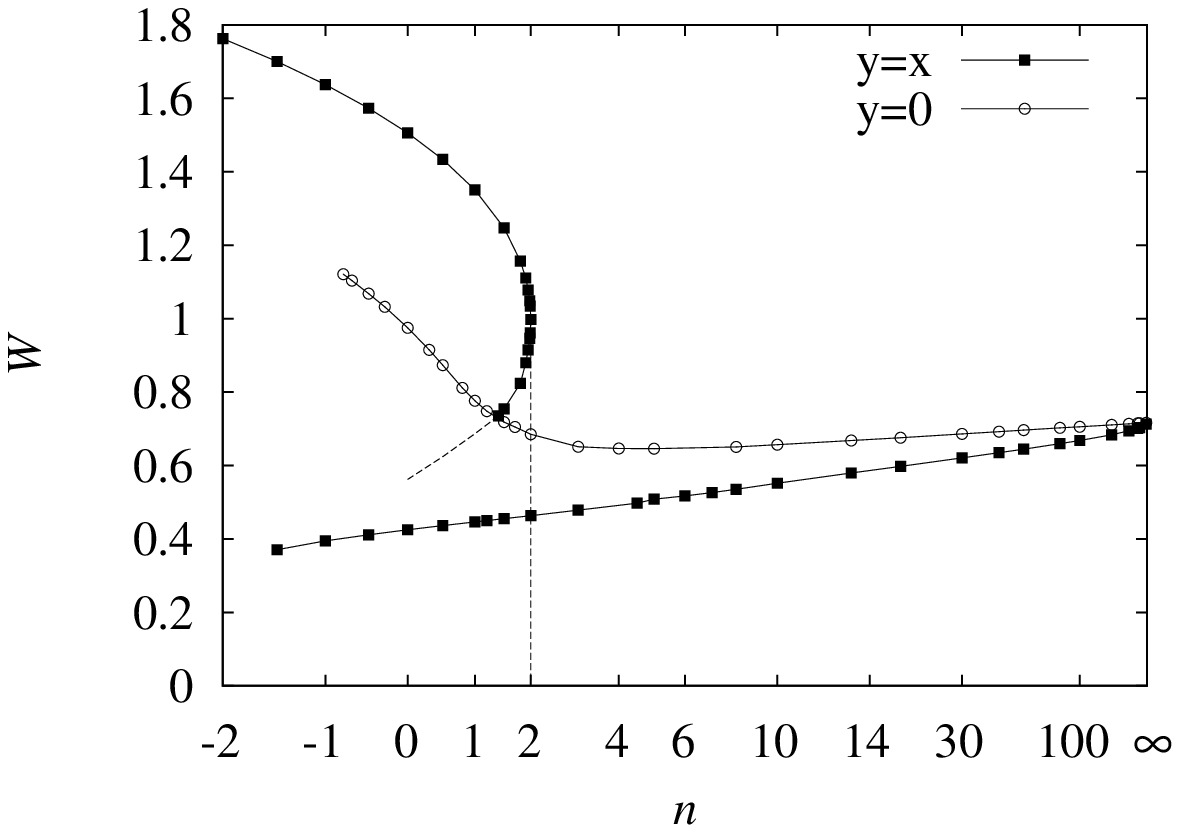}
\caption{
The phase diagram of the O($n$) loop model on the square lattice in the
($n,x$) parameter space. 
In order to map the range $-2<n<\infty$ on a finite interval,
a $1-8/(n+10)$ scale is chosen along the horizontal axis. The vertical
axis shows the temperature-like quantity $W$ defined in the text. 
The curved line on the left is the O($n$) critical
line, and a part of its continuation into the low-temperature O($n$)
phase which exists only for $n\leq 2$. The vertical dashed line shows
the boundary of this phase at $n=2$. The line of lattice-gas-like
transitions extends all the way to $n=\infty$, where it ends in a point
that is accurately known as described in the text.}
\label{phdiag}
\end{figure}

The phase diagram for $n$ in the range $(-2, \infty)$ is shown in
Fig.~\ref{phdiag}.
In order to map the range $-2<n<\infty$ on a finite interval,
a $1-8/(n+10)$ scale is chosen along the horizontal axis. The vertical
axis shows the temperature-like quantity $W\equiv 1/[x(n+10)^{1/4}]$,
which parametrizes the bond weight, while remaining finite in the
mentioned interval. The curved line on the left is the O($n$) critical
line, and a part of its continuation into the low-temperature O($n$)
phase which exists only for $n\leq 2$.

\subsection{subspace $y=0$, $z=x^2$}
\label{y0sec}
For $y=0$, there exists an exactly solvable case $x_c=z_c=1/2$, 
which is called branch 0 \cite{BN}. For $n=0$ it describes the $\theta$
point of a collapsing polymer \cite{BBN}.  For other values of $n$,
it describes a higher critical point (but not the tricritical point
analyzed in Ref.~\cite{GNB}). Since the present value $z=x^2$ is smaller
than that for branch 0, we do not expect that the $y=0$, $z=x^2$
subspace contains the $n=0$ collapse transition.
However, the fact that an Ising degree of freedom is associated with each
separate loop implies that a degree of Ising ordering is introduced
at the critical points for $n<2$, where the largest loops are expected
to diverge. Thus, we may expect that, at least for some values of $n$,
transitions occur in a different universality class than that of branch 1.
Furthermore it remains to be investigated if the phase diagram displays
the same topology as that for the $y \ne 0$ case. 

First, we investigate that the lattice-gas-like transition persists in the
present subspace with $y=0$.  
For $n \ge 8$, we solved for $x_c$ in the scaling equation Eq.~(\ref{xhs2}),
using $X_s(L)=1/8$ as the Ising magnetic scaling dimension. 
For $n=2,3,4,5$, we solved instead the scaling equation Eq.~(\ref{xhs}) for
$X_s(L)$. Only even $L$ are used in these calculations.
We found that the solutions converge in the way described by 
Eq.~(\ref{xcs}), confirming that Eq.~(\ref{xhse}) applies to $X_s(L)$,
thus indicating that algebraic decaying of the staggered correlation 
function occurs in the thermodynamic limit. 
After extrapolation of the critical points $x_c$,
we calculated the free energy density $f(L)$ at $x_c$ for even system sizes up
to $L=18$. A fit of the data according to Eq.~(\ref{fc}) then yields
the conformal anomaly $c$. The results behave in a way similar to the
$y=x$ case: in accurate agreement with Ising universality ($c=1/2$), 
except for a few relatively small $n$ values. 
The numerical results for $x_c$ and $c$ are included in Table \ref{numdx}.

Next we address the question whether the critical manifold continues into 
the $n \le 2$ range and connects to an Ising-dense O($n$)  transition.
We handle this problem by solving $x_c(L)$ in
the scaling equation (\ref{xhs}), for even system sizes up to $L=18$. 
For $n<1$, the magnetic scaled gap $X_h(L)$ is used in Eq.~(\ref{xhs}). 
We find that the solution $x_c(L)$ converges with $L$
in the way described by Eq.~(\ref{xcs}). 
The estimated critical points are included in the phase diagram in the
$(x,n)$ plane shown in Fig. \ref{phdiag}.
We then calculated the magnetic scaled gaps $X_h(L)$ for a sequence of
systems with even $L$ up to $18$ at the solutions $x_{c}(L)$.
Extrapolation of the gaps according to Eq.~(\ref{xis}) yields the scaling
dimension $X_h$, which is listed in Table \ref{numdxnle2c}.
These results are, in a limited range, compatible with the known
scaling dimension $X_h$  of the critical O($n$) transition, but the
accuracy is low because of strong corrections to scaling, with the
exception of the result at $n=0$.

We also calculated the free energy density at the estimated $x_c$.
A fit of these data by Eq.~(\ref{fc}) then yields estimates of the
conformal anomaly $c$ for this transition, which are also listed in
Table \ref{numdxnle2c}. For most values of $n$ these results do not
agree with the known theory for the O($n$) critical line, or with a 
superposition of O($n$) criticality and Ising behavior.

For small finite systems with $n=-1$, the leading eigenvalue $\Lambda_0$
becomes twofold degenerate at $x=1/2$. The same applies to the leading
eigenvalue $\Lambda_1$ in the odd (magnetic) sector.  Moreover, these two 
pairs of eigenvalues are also equal. On this basis we
conjecture that $x_c(-1)=1/2$ and $X_h(-1)=0$.

For $n \ge 1$, we found no solutions of the scaling equation
Eq.~(\ref{xhs}) for $X_h(L)$ with even $L$.  The staggered scaled gap 
$X_s(L)$ was
used instead to study the possible transition. 
The scaling equation was solved for a sequence of even systems up to $L=18$.
We find that the solutions $x_c(L)$ behave in a way consistent with
convergence to a critical point $x_c$ as described in Eq.~(\ref{xcs}). 
The estimated critical points are included in the phase diagram in the
$(x,n)$ plane shown in Fig. \ref{phdiag}.

Next we calculated the scaled staggered gaps $X_s(L)$ for even system sizes
up to $L=18$ at the solutions $x_{c}(L)$. The gaps converge to the
scaling dimension $X_s$ according to Eq.~(\ref{xis}). 
Unfortunately, the convergence of the data is not good except for $n=1$.
The results for
$X_s$ are listed in Table \ref{numdxnle2c}. For $n=1$ numerical result 
for  $x_{c}$ agrees with the self-dual value $x=1/\sqrt{2}$ (see
Sec.~\ref{disc} for further details), and the latter value was used to 
estimate the universal quantities for $n=1$. 

We also calculated the free energy density for $1 \leq n \leq 2$ at the
estimated critical bond weight $x_c$.
>From a fit of  Eq.~(\ref{fc}) to these data, we estimate the conformal
anomaly $c$ for this transition, as also listed in Table \ref{numdxnle2c}.

\begin{table}[tpbd]
\caption{Numerical results for the critical points $x_c(n)$, the
conformal anomaly $c(n)$ and scaling dimensions $X_h(n)$ and $X_s(L)$ 
for several values of $n$ in the subspace $y=0$. Since solutions of 
Eq.~(\ref{xhs}) for $X_h$ are absent for $n\ge 1$, the result $X_h(1)$ was
obtained from a fit of the $X_{h,L}$ data at the self-dual value
$x=1/\sqrt{2}$ of the bond weight. The latter value was also used to
estimate $X_s(1)$ and $c(1)$. It is in accurate agreement with the
numerical result obtained by fits to the $X_s(L)$ data for $n=1$.
Satisfactory finite-size convergence is found only for $n=-1$, 0 and 1.
The error margins, shown between parentheses, are difficult to estimate
in some cases indicated with a question mark. 
}
\begin{tabular}{|c|c|c|c|}
\hline
$n$     & $x_c(n)$      & $X_h(n)$    &$c(n)$     \\ 
\hline
$-1.0$  & 1/2           & 0           &$-$2.00(1) \\
$-0.8$  & 0.51229(1)    & 0.06(2)?    &$-$1.115(2)\\
$-0.7$  & 0.51889(1)    & 0.07(2)?    &$-$0.92(1) \\
$-0.5$  & 0.53317(2)    & 0.07(1)     &$-$0.60(1) \\
$-0.3$  & 0.54910(1)    & 0.09(2)     &$-$0.339(1)\\
0.0     & 0.57686(2)    & 0.10417(5)  & 0         \\
0.3     & 0.6106(3)     & 0.10(3)     & 0.306(1)  \\
0.5     & 0.637(2)      & 0.16(4)?    & 0.506(1)  \\
0.8     & 0.68(1)       & 0.21(3)?    & 0.81(1)   \\
1.0     & $1/\sqrt{2}$  &0.25000000(5)&1.0000 (1) \\
\hline
$n$     & $x_c(n)$      &$X_s(n)$     &$c(n)$     \\
\hline
1.0     & $1/\sqrt{2}$  &1.00000000(1)& 1.0000 (1)\\
1.2     & 0.731(2)      & 0.81(4)     & 1.13(2)   \\
1.5     & 0.755(4)      & 0.5(2)      & 1.32(1)   \\
1.7     & 0.768(2)      & 0.3(3)      & 1.41(1)   \\
2.0     & 0.784(1)      & 0.1(3)      & 1.49(2)   \\
\hline
\end{tabular}
\label{numdxnle2c}
\end{table}

\section{Discussion}
\label{disc}
Using a finite-size-scaling analysis of results from transfer-matrix
calculations,  we have determined the phase diagram of the O($n$) loop
model on the square lattice in the  $(x,n)$ plane, where $x$ is the
weight of a lattice edge covered by a loop. Two subspaces, $y=x$ and
$y=0$, were investigated.

For $n>>2$ we find an Ising-like phase transition associated with
the onset of a checkerboard-like ordering of the elementary loops.
In this respect, the ordered state resembles that
of the hard-square lattice gas. For large values of $n$ the critical
points shown in Fig.~\ref{phdiag} approach the accurately known
lattice-gas limit. For the case $y=x$ the data in this figure suggest
that this approach happens with a weak  cusp-like singularity.
This behavior can be explained by the residual presence of loops
exceeding the size of an elementary faces. The next-smallest loops
cover a rectangle with the size of two faces, and contain two $y$-type
vertices, at the expense of an extra  weight factor $x^2$. The presence
of these larger loops thus corresponds with a repulsive potential of
order $x^2 \propto 1/\sqrt{n}$ between next-nearest-neighboring hard
squares. In lowest order one then expects a linear dependence of the
critical chemical potential of the hard-square model on such a repulsion.
Noting that the quantity $W$ shown in Fig.~\ref{phdiag} plays the
role of this chemical potential, one expects that the critical value 
of $W$ depends linearly on $1/\sqrt{n}$ for large $n$. This   
corresponds with a square-root like singularity on the scale used in
Fig.~\ref{phdiag}, which behaves as $1/n$ for large $n$. This explains
the weak  cusp-like singularity.

The appearance of loops exceeding the size of an elementary square,
in particular loops covering two squares, can be interpreted as a
softening of the nearest-neighbor repulsion. In this respect, the loop
model with $n<\infty$ approaches the experimental situation of monatomic
gases adsorbed on the (1,0,0) surface of a cubic crystal better than the
hard-square model.

These results are in part similar to those obtained for the large-$n$
loop model on the honeycomb lattice \cite{ngt2tr}, which behaves as
a hard-hexagon model \cite{Baxhh}, with a phase transition in the
three-state Potts universality class. It appears that large-$n$ loop
models generically approach the behavior of systems of hard particles,
with universal properties that are dependent on the microscopic lattice
structure. The universal properties may also depend on the allowed
set of vertices. For instance, if we put the vertex weights $y=z=0$
in the loop model on the square lattice, then the corresponding
hard-square model is subject not only to nearest-neighbor exclusion,
but also to next-nearest-neighbor exclusion. The universal behavior
of this system is not Ising-like \cite{FBN,KR}.

The question about a possible physical interpretation of this
large-$n$ transition of the loop model in the language of the spin
model specified by Eq.~(\ref{Zspin}) is answered by substitution of
the numerical results $x_c$ for the critical point, and the length
scale $1/\sqrt n$ of the spin vectors in that equation. This shows that
the Boltzmann factors of the spin model with $n>2$ can become negative
at the phase transition. This exposes the unphysical nature of the 
lattice-gas-like transition in the language of the spin model.

In Sec.~\ref{results} we have also investigated the critical properties
of this O($n$) model in the range $-2\leq n\leq 2$. 
In the case $x=y$, we found that the lattice-gas-like transition
line continues into the low-temperature O($n$) phase.
The universal behavior along this part of the transition
line is interpreted as a superposition of Ising criticality
and dense O($n$) loop model behavior, similar to earlier findings for a
related square-lattice O($n$) model. For the special point $n=1$, 
the O($n$) critical point and the Ising-like transition are dual
images of one another. The duality transformation includes the sum
on the weights of the two $z$-type vertices, thus leading to a single
4-leg vertex with weight $2z$, and a replacement of the
loop segments by empty edges and vice versa. The transformation
maps the $x$- and $y$-type vertices on the same type, and interchanges
the empty and the $z$-type vertices. The normalization of the weight of
the empty vertex to 1 thus reduces $x$ and $y$ by a factor $2z$, and
changes $z$ into $1/4z$. The $n=1$ results for the
critical points $x_c$ found in Tables \ref{yne0nlt2} and \ref{numdxlt}
satisfy this dual relation with one another.

Also for the case $y=0$, we find that the lattice-gas-like transition line
continues into the range $n<2$, but the topology of the ($x,n$) phase
diagram is different.  It does not enter into the dense O($n$) phase,
but continues as a line of critical points separating the disordered 
phase from the dense O($n$) phase. Our numerical data suggest that its
universal properties do not match those of the critical O($n$) line
(for most of the range $n<1$), or those of a superposition of Ising and
critical O($n$) behavior. Furthermore, our results for the universal
quantities differ significantly from those reported for branch 3 of the
square O($n$) model \cite{BN,BNW} and for the tricritical O($n$)
model \cite{NGB}, at least for most of the range $n<1$. We remark that 
both branch 3 of the square O($n$) model and our $y=0$ critical
line lie relatively close to the higher critical point of branch 0
($x=z=1/2,~ y=0$), but for $-1<n<1$ they reside on different sides of
branch 0. This allows for the possibility that the $y=0$ critical point
although also twofold unstable, is attracted by a different fixed point
than branch 3. Unfortunately, the limited accuracy of our numerical
data for the $y=0$ line case impedes the further identification of its
universal nature in terms of possibly existing exact results.

As already mentioned in Sec.~\ref{intro}, there exists an Ising-like
degree of freedom that is frozen out along each loop for $y=0$. Thus,
for $n=0$, where we allow only a single loop, it plays no role and
indeed we find the O(0) or SAW critical behavior. For $n \ne 0$ 
more loops may appear, whose Ising degree of freedom may differ. 
But, depending on the weight $z$, neighboring loops will tend to
meet at $z$-type vertices, and thus assume the same Ising variable.
It may thus be expected that 
at the $y=0$ O($n$) critical point, where the largest loop diverges,
there will also be some ordering of the Ising degrees of freedom,
thus allowing universal behavior that is different from that of the
generic O($n$) critical point.

For $n=1$, the numerical result for $x_c$ agrees, within an error margin
of about $10^{-5}$, with the self-dual location $x_c(1)=1/\sqrt 2$.
Furthermore, this self-dual point can be mapped on Baxter's 8-vertex
model \cite{Baxb}, by adding down- or left-pointing arrows on the edges
covered by an O($n$)  loop, and up- or right-pointing arrows to the
empty edges. Since two of the vertex weights are zero, the symmetry
relations of the 8-vertex model allow a further mapping on the 6-vertex
model, with vertex weights $(a,b,c)=(1/\sqrt {2},1/\sqrt {2},1)$ in the
notation used in Ref.~\cite{Baxb}.

Also the conjectured critical point $x_c(-1)=1/2$ mentioned in
Sec.~\ref{y0sec} can be given a more firm basis. We recall that the weight
$z$ is actually redundant for $n=-1$ \cite{BN}. To see this, consider an
arbitrary loop configuration where four loop segments come in at a
given vertex. There are two possible ways to connect these segments by
a $z$-type vertex, and the numbers of loops closed differ by precisely
1. Taking into account that the loop weight is $-1$, one observes that 
the two contributions due to the summation on the two $z$-type vertices
cancel. Thus, in effect, configurations with $z$-type vertices do not
contribute to the partition sum. Therefore, the $n=-1$ point of branch 0
in Ref.~\cite{BN}, namely $x=1/2$, is an exact critical point in our
$y=0$ subspace, in spite of the fact that the weight $z$ is different.
The results $X_h=0$ and $c=-2$ found there agree with the present findings.

\acknowledgments
We are much indebted to Prof. B. Nienhuis for freely sharing his insights
in various subtleties of O($n$) loop models.
Z.~F. acknowledges hospitality extended to her by the Lorentz Institute.
This work was supported by the NSFC under Grant No.~11175018, and 
by the Lorentz Fund.

\end{document}